\documentclass[preprint,showpacs,preprintnumbers,amsmath,amssymb,superscriptaddress]{revtex4}

\include{ams}
\usepackage{amsmath,amssymb,amsthm}
\usepackage{dcolumn}
\usepackage{bm}
\usepackage{graphicx}

\begin{document}

\title{Electronic pathway in the photosynthetic reaction centers and some mutation of RC's}

%
%

\author{M. Pudlak}
\email{pudlak@saske.sk}
\affiliation{Institute of Experimental Physics, Slovak Academy of Sciences, Watsonova 47,043 53 Kosice, Slovak Republic}

\author{R. Pincak}
\email{pincak@saske.sk} \affiliation{Institute of Experimental
Physics, Slovak Academy of Sciences, Watsonova 47,043 53 Kosice,
Slovak Republic} \affiliation{Joint Institute for Nuclear Research,
BLTP, 141980 Dubna, Moscow region, Russia}




\date{\today}

\begin{abstract}

The reaction center of {\it Chloroflexus aurantiacus} and {\it
Rhodobacter sphaeroides} mutation of RC`s was investigated. To
describe the kinetic of the {\it Chloroflexus aurantiacus} RC's we
use incoherent model of electron transfer. It was shown that the
asymmetry in electronic coupling must be included to explain the
experiments. For the description of {\it Rhodobacter sphaeroides}
{\it H(M182)L} mutation we used partially coherent as incoherent
models of electron transfer. These two models are discussed with
regard to the observed electron transfer kinetics. It can be
concluded that partially coherent model is more adequate. We
predict some new electron pathways for describing the kinetic of
RC's and some mutation.

\end{abstract}

\maketitle

\section{Introduction}

Photosynthesis is a reaction in which light energy is converted into
chemical energy. The primary process of photosynthesis is carried
out by a pigment-protein complex embedded in the membrane, that is,
RC. In photosynthetic purple bacteria, the cyclic electron transfer
reaction is performed by RC and two other components: the cytochrome
(Cyt) $bc_1$ complex, and the soluble electron carrier protein. The
photosynthetic reaction centers (RC) is a special pigment-protein
complex, that functions as a photochemical trap. The precise details
of the charge separations reactions and subsequent dark electron
transport (ET) form the central question of the conversion of solar
energy into the usable chemical energy of photosynthetic organism.
The function of the reaction center is to convert solar energy into
biochemical amenable energy. Therefore, we wish to understand which
features of the reaction center are responsible for the rate
constants of these reactions.

Insight into the molecular organization of the RC has been
derived, initially, from spectroscopic studies and, subsequently,
from the development and analysis of high-resolution crystal
structures of several photosynthetic organisms. The first RC
structurally resolved (3 {\textmd{\AA}}) was of the purple
bacterial RC from {\it Rhodopseudomonas viridis}~\cite{Deisen}.
This was soon followed by the elucidation of several other purple
bacterial structures. Good progress is also being made toward
achieving two- and three-dimensional structures of photosystem II
(PSII) crystals. It is surprising that the structures of all of
the different RC's show a dimeric core with a pseudo-$C_2$ axis of
symmetry.

A remarkable aspect of the RC structures is the occurrence of two
almost identical electron acceptor pathways arranged along the $C_2$
axis relative to the primary charge-separating dimer (bacterio)
chlorophyll (Fig.~1). This finding posed a key question: Does
electron transfer involve both branches? In the purple bacterial RC,
only one branch is active although the inactive branch can be forced
into operation with modification of amino acid side chains on the
active branch~\cite{Allen}. Moreover the charge-separating electron
transfer reactions occur with a remarkably high quantum yield of
$96$\%, where from two possible symmetric branches only the branch
$L$ is active in the electron transfer. This efficiency relies on
the rates of the charge-separating reactions being 2-3 orders of
magnitude faster than the rates of the competing reactions.

The strong asymmetry imposed on primary charge separation
photo-chemistry in the purple bacterial RC results from two
homologous polypeptides that function as a heterodimer. A
heterodimer is also involved in the core of the RC's of PSI and
PSII. However, some RC's, such as heliobacteria~\cite{Amesz} and
green sulfur bacteria~\cite{Sakurai}, contain two identical
homodimeric polypeptides, and electron transfer is potentially
bifurcated.

Genetic sequence information has greatly improved the understanding
of the origin of the RC proteins. From the sequence analysis, it
became clear that the purple bacteria RC is remarkably similar to
that of PSII, and PSI was also discovered to have similarity with
that of the green sulfur bacteria~\cite{Golbeck}. Recent structural
comparisons between PSI and PSII, for example, show a distinct
structural homology, which suggests that even these two RC's likely
share a common ancestor~\cite{Schubert}.

In purple bacteria, the electron rate is sensitive to the free
energy difference between the excited state and the charge-separated
state but not to the relative distribution of electrons over the two
macrocycles of the donor. After extensive studies, the rate is now
established to be critically coupled to the properties of the
bacteriochlorophyll monomer that lies between the donor and
bacteriopheophytin acceptor (Fig.1). The involvement of the
bacteriochlorophyll monomer may give rise to multiple pathways for
electron transfer~\cite{Van} and can partially determine the
asymmetry of the electron transfer along one branch~\cite{Heller}.
We believe that the reason for asymmetric ET between prosthetic
groups located on different polypeptides is a different molecular
dynamics. Dynamics of atoms causes the change of the electrical
potential fields and the conformational variations influence the
mutual orientations between cofactors. Then the energy gap and
overlap of electronic wave functions fluctuates as a result in the
system. The net result is a different fluctuation of electronic
energy levels on prosthetic groups and also a different fluctuations
of the overlaps of the electronic wave functions on $L$ and $M$
branches. On the other hand the chain located on subunit $M$ is
inactive in ET and the highly asymmetric functionality, however, can
be decreased by amino acid mutations or cofactor modification. We
used this approach to explain the effect of individual amino acid
mutation or cofactor modifications on the observed balance between
the forward ET reaction on the $L$-side of the RC, the charge
recombination processes, and ET to the $M$-side
chromophores~\cite{Kirmaier,Takahashi,Shuvalov,Gehlen,Holzwarth0}.

The theoretical models that describe the charge transfer in
reaction centers  using parameters with clear physical
interpretations. Some of these input parameters can not be deduced
from independent experimental work. The information regarding the
energetic parameters, the medium reorganization energies, the high
frequency modes, and electronic coupling terms can be achieve with
quantum mechanical computations. But until now these parameters
which characterize the reaction centers are not available. And so
we use the set of parameters which fit the experiments. Several
sets of parameters were used to describe a charge transfer in the
RC. A set of parameters based on molecular dynamics
simulations~\cite{Marchi} corresponds to a dominance of
superexchange mechanisms for the primary ET reaction in RC's.
Another set of parameters~\cite{Tanaka,Bixon} was used to fit
experimental data. This second set of parameters derives a
dominant contribution from the sequential mechanism. The first set
of parameters has the larger reorganization energies and the
greater coupling factors. This set of parameters makes the ET rate
much larger than it is found in the wild-type proteins. The
possibility to find out the input parameters from theory is
comparison of observed kinetics for different mutated reaction
centers. The problem is that not always the impact of mutation on
the input parameters is clear. In this paper we focus on the
electron transfer in two RCs. First is the RCs of the green
bacterium  {\it Chloroflexus aurantiacus}. The second is the RCs
of {\it Rhodobacter sphaeroides} {\it H(M182)L} mutation. It is
believe that both purple bacterial RCs and RCs from {\it C.
aurantiacus} have a similar structure~\cite{Holzwarth}. We adapt
in this work the set of parameters that characterized the observed
L-side experimental kinetics of wild-type (WT) RCs of
Rb.sphaeroides very well. The {\it Chloroflexus aurantiacus} RC's
and the RCs of {\it Rhodobacter sphaeroides} {\it H(M182)L}
mutation have structural similarity but charge separation kinetics
are different. Both these RCs contain BPheo pigment in M-branch in
the position where BChl monomer is placed in the WT reaction
center. In contrast with this structural similarity, the H(M182)L
mutant reveal the electron transfer through the M branch, in the
{\it Chloroflexus aurantiacus} RC's the M-branch is inactive.

\section{Theory} We attempt to analyze the possibility that ET
asymmetry can be described by model which assumes that there exists
the vibrational modes of the medium which has a sufficient time for
relax to the thermal equilibrium after each ET step. We start by
considering an electron transfer system in which the electron has
$N$ accessible sites, embedded in a medium. We denote by $|j\rangle$
the state with electron localized at the $j$th site and
$j=1,2,...,N$. The $j$ and $k$ sites are coupled by $V_{jk}$. The
interaction of the solvent with the system depends on the electronic
states $|j\rangle$ by $H_j$. The total model Hamiltonian for the
system and medium is
\begin{eqnarray}\label{} H=H_0+V ,
\end{eqnarray} where
\begin{eqnarray}
\label{}
H_0=\sum_{j=1}^{N}|j\rangle[\varepsilon_j-i\Gamma_j+H_j]\langle j|,
\end{eqnarray}
\begin{eqnarray} \label{} V=\sum_{j,k=1}^{N}V_{jk}|j\rangle \langle
k|,\quad j\neq k,
\end{eqnarray}
where $\varepsilon_j$ is the site
energy. The parameter $\hbar/2\Gamma_j$ has a meaning of the
lifetime of the electron at site $j$ in the limit of the zero
coupling parameter. It can characterize the possibility of the
electron escape from the system by another channel, for instance a
nonradiative internal conversion or recombination process.

The Hamiltonian describing the reservoir consisting of harmonic
oscillators is
\begin{eqnarray} \label{}
H_j=\sum_a\bigg\{\frac{p_\alpha^{2}}{2m_\alpha}+\frac{1}{2}m_\alpha\omega_\alpha^{2}(x_\alpha-d_{j\alpha})^{2}\bigg\}.
\end{eqnarray}Here, $m_\alpha$ and $\omega_\alpha$ are frequency and the mass
of the $\alpha$th oscillator, and $d_{j\alpha}$ is the equilibrium
configuration of the $\alpha$th oscillator when the system is in the
electronic state $|j\rangle$. The total density matrix $\rho(t)$ of
the ET system and the medium satisfies the Liouville equation,
\begin{eqnarray}
\label{} \partial_t\rho (t)=-\frac{i}{\hbar}[H \rho(t)-\rho(t)
H^{\dagger}]=-i L\rho(t). \end{eqnarray} In the interacting picture,
\begin{eqnarray}\label{3th6} \rho_I(t)=\exp\bigg( \frac{i}{\hbar}H_0 t
\bigg)\rho(t)\exp\bigg( -\frac{i}{\hbar}H_0^{\dagger}t \bigg).
\end{eqnarray} The Liouville equation in the interacting picture has
the following form:
\begin{eqnarray} \label{}
\partial_t\rho (t)=-\frac{i}{\hbar}[V_I(t) \rho_I(t)-\rho_I(t)
V_I^{\dagger}(t)]=-i L(t)\rho_I(t),
\end{eqnarray}where
\begin{eqnarray}\label{} V_I(t)=\exp\bigg( \frac{i}{\hbar}H_0 t
\bigg)V\exp\bigg( -\frac{i}{\hbar}H_0 t \bigg). \end{eqnarray} Here
we denote the total trace, and the partial traces over the ET system
and over the medium by $Tr$, $Tr^{e}$, $Tr^{Q}$, respectively. By
definition $Tr\equiv Tr^{Q}Tr^{e}$. The population on state
$|j\rangle$ at time $t$ is given by
\begin{eqnarray} \label{}
P_j(t)=Tr(|j\rangle\langle j|\rho(t)). \end{eqnarray} We assume that
the vibrational relaxation is sufficiently rapid so that the system
can relax to thermal equilibrium after each ET step. This assumption
determines a choice of projector operator. The projector operator
$D$ acting on an arbitrary operator $B$ in the Hilbert space of the
total ET system and medium is defined by~\cite{Mukamel}
\begin{eqnarray} \label{}
DB=\sum_{j=1}^{N}Tr(|j\rangle\langle j|B)\rho_j |j\rangle\langle j|,
\end{eqnarray} where $\rho_j$ is the equilibrium medium density
matrix in the state $|j\rangle$, i.e.,
\begin{eqnarray} \label{}
\rho_j=\frac{\exp(-H_j/k_BT)}{Tr^{Q}\exp(-H_j/k_BT)}.
\end{eqnarray}
Using the standard projection operator
techniques~\cite{Zwanzig,Shibata} we can derive a generalized
master equation for the populations,
\begin{eqnarray} \label{}
\partial_tP_j(t)&=&-\frac{2\Gamma_j}{\hbar}P_j(t)-\sum_{k=1}^{N}\int_0^{t}W_{jk}(t-\tau)P_j(\tau)d\tau\nonumber\\[1mm]
& & +\sum_{k=1}^{N}\int_0^{t}W_{kj}(t-\tau)P_k(\tau)d\tau,\quad
j=1,...,N,\quad j\neq k, \nonumber\\[1mm]
& & \end{eqnarray} where
\begin{eqnarray} \label{3th14}
W_{jk}(t)&=&2\frac{|V_{jk}|^{2}}{\hbar^{2}}\textrm{Re}\bigg\{\exp\bigg[-\frac{\Gamma_j+\Gamma_k}{\hbar}t\bigg]\exp\bigg[\frac{i(\varepsilon_j-\varepsilon_k)}{\hbar}t\bigg]\nonumber\\[1mm]
& & \times
\exp\bigg\{\sum_\alpha\frac{E_{jk}^{\alpha}}{\hbar\omega_\alpha}[(\bar{n}_\alpha+1)e^{-i\omega_\alpha
t}+\bar{n}_\alpha e^{i\omega_\alpha t}-(2\bar{n}_\alpha+1)]\bigg\}\bigg\}.\nonumber\\[1mm]
& & \end{eqnarray} Here, $
\bar{n}_\alpha=[\exp(\hbar\omega_\alpha/k_BT)-1]^{-1}$ is a thermal
population of the $\alpha$th mode and
\begin{eqnarray} \label{}
E_{jk}^{\alpha}=\frac{1}{2}m_\alpha\omega_\alpha^{2}(d_{j\alpha}-d_{k\alpha})^{2}
\end{eqnarray} is the reorganization energy of the $\alpha$th mode
when system transfer from state $|j\rangle$ to state $|k\rangle$.

\section{Model of Reaction Center}
\label{nm}

To describe the first step of electron transfer processes in the
reaction centers we have used the 5-sites kinetic model of RC.

\begin{figure}[htb]
\centering
\includegraphics[width=0.5\textwidth]{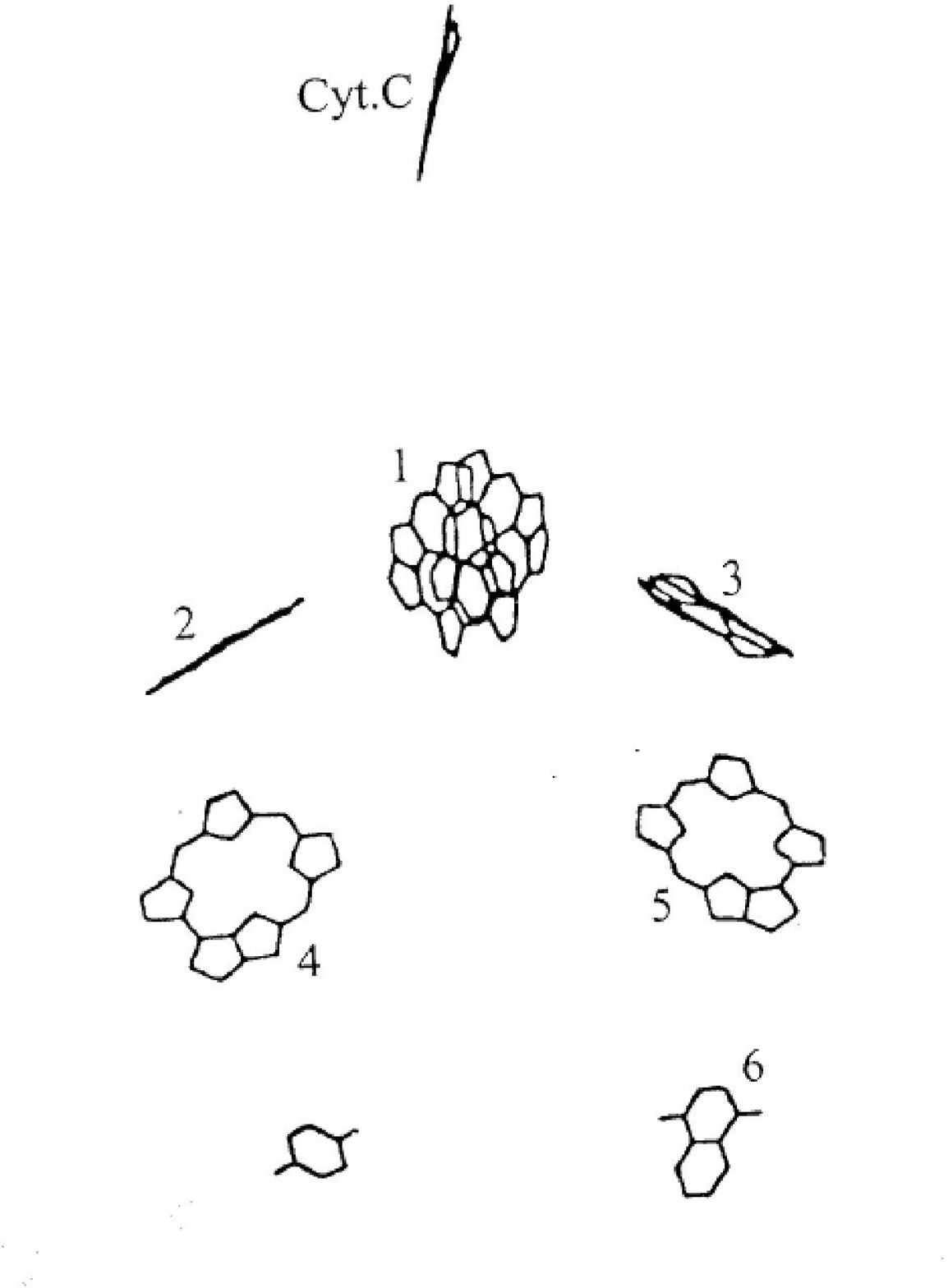}
\caption{ The RC of purple bacteria are composed of three protein
subunits called $L$, $M$ and $H$. Dimer $P$ is describing by
molecule 1. Cofactors in the subunits L are: 3 represent
($\mathrm{BChl}_L$) molecule 5 ($\mathrm{BPh}_L$) and 6 is
($\mathrm{Q}_L$) and identically in the subunits $M$
($\mathrm{BChl}_M$) is describing by molecule 2 and molecule 4
represent ($\mathrm{BPh}_M$). Cytochrom C serve as a source of
electrons for reaction center.}\label{kin1}
\end{figure}

We designate the special pair $P$ as site $1$, the sites $2$ and
$3$ represent the molecules $\mathrm{BChl}_M$ and
$\mathrm{BChl}_L$, and the sites $4$ and $5$ then represent the
molecules $\mathrm{BPh}_M$ and $\mathrm{BPh}_L$ (Fig.~1). We
assume that we can neglect the backward electron transfer from
quinone molecules and so we use the complex energies of  4,5
molecules of RC. Based on experimental observations of ET in RC,
it is expected that bacteriochlorophyll play a crucial role in ET.
In this 5-sites model we have assumed that ET in RC is sequential
where $\mathrm{P^+BChl^-}$ is a real chemical intermediate. The
imaginary part of energy level 1 describes the probability of
electron deactivation to the ground state. We describe the ET in
the {\it Chloroflexus aurantiacus} RC's by the following kinetic
model


\begin{subequations} \begin{eqnarray}
\partial_tP_1(t)&=&-(\frac{2\Gamma_1}{\hbar}+k_{12}+k_{13})P_1(t)\nonumber\\[1mm]
& & +k_{21}P_2(t) + k_{31}P_3(t),\\[1mm]
\partial_tP_2(t)&=& -(k_{21}+k_{24})P_2(t)+k_{12}P_1(t)+k_{42}P_4(t),\\[1mm]
\partial_tP_3(t)&=&
-(k_{35}+k_{31})P_3(t)+k_{13}P_1(t)+k_{53}P_5(t),\\[1mm]
\partial_tP_4(t)&=&
-(\frac{2\Gamma_M}{\hbar}+k_{42})P_4(t)+k_{24}P_2(t),\\[1mm]
\partial_tP_5(t)&=& -(\frac{2\Gamma_L}{\hbar}+k_{53})P_5(t)+k_{35}P_3(t). \end{eqnarray}
\end{subequations}
Here we denote $k_{ij}(s\rightarrow 0^{+})=k_{ij}$ and
$k_{ij}(s\rightarrow 0^{+})=\int_{0}^{\infty}W_{ij}(t)dt $.

We assume that the rate constant which characterizes ET can be
described by both a low frequency medium vibrational mode and a
high frequency intramolecular vibrational mode. At a high
temperature regime the constant $k_{ij}(s\rightarrow0^{+}) $ is in
the form~\cite{Jortner}:
\begin{equation}\label{}
\begin{split}
k_{ij}&=\int_{0}^{\infty}W_{ij}(t)dt=\frac{2\pi}{\hbar}V_{ij}^{2}(\frac{1}{4\pi\lambda_{mij}k_{B}T})^{1/2}\exp(-S_{cij})\\
&\quad\times\sum^{\infty}_{n=0}\frac{S_{cij}^n}{n!}\exp[-\frac{(G_{ji}+\lambda_{mij}+n\hbar\omega_{cij})^2}{4\lambda_{mij}k_{B}T}].
\end{split}\end{equation}\\

Here, \( G_{ij}=\epsilon_{i}-\epsilon_{j} \) and \(
S_{cij}=\frac{1}{2\hbar}m_{cij}\omega_{cij}(d_{ci}-d_{cj})^2 \) is
the scaled reorganization constant for the high frequency $ij$-th
mode, which is nonzero when electron is transferring from the
state $|i\rangle$ to the state $|j\rangle$, and \(
\lambda_{mij}=\frac{1}{2}m_{mij}\omega_{mij}^2(d_{mi}-d_{mj})^2 \)
is the reorganization energy of the low-frequency mode when the
electron is transferring from the state $|i\rangle$ to the state
$|j\rangle$. The back electron transfer reaction rate constant can
be calculated by using the detailed balance relation and can be
expressed in the form $k_{ji}=k_{ij}\exp(-G_{ij}/k_{B}T)$.

The quantum yields $\Phi_L$, $\Phi_M $ of electronic escape via
branch $L$, $M$ and the quantum yields $\Phi_G$ of direct ground
state recombination can be characterized for 5-sites sequential
kinetic model by the expressions
\begin{subequations}
\begin{align}\label{}
{\phi_G}& ={\frac{2\Gamma_1}{\hbar}P_{1}(s\rightarrow0^{+})},\\
{\phi_L}& ={\frac{2\Gamma_L}{\hbar}P_{5}(s\rightarrow0^{+})},\\
{\phi_M}& ={\frac{2\Gamma_M}{\hbar}P_{4}(s\rightarrow0^{+})},
\end{align}\end{subequations}\\
where the expression $\Phi_L+\Phi_M +\Phi_G=1$ have to by
fulfilled. The analytical expressions for the ratio of the quantum
yields have the forms
\begin{subequations}
\begin{align} \label{}
\frac{{\phi_L}}{{\phi_M}}&=\frac{k_{13}k_{35}(\frac{2\Gamma_M}{\hbar}k_{24}+k_{21}(k_{42}+\frac{2\Gamma_M}{\hbar}))\frac{2\Gamma_L}{\hbar}}
{k_{12}k_{24}(\frac{2\Gamma_L}{\hbar}k_{35}+k_{31}(k_{53}+\frac{2\Gamma_L}{\hbar}))\frac{2\Gamma_M}{\hbar}},\\
\frac{{\phi_L}}{{\phi_G}}&=\frac{k_{13}k_{35}\frac{2\Gamma_L}{\hbar}}
{(\frac{2\Gamma_L}{\hbar}k_{35}+k_{31}(k_{53}+\frac{2\Gamma_L}{\hbar}))\frac{2\Gamma_1}{\hbar}}.\end{align}\end{subequations}

The results of numerical calculations of QY's rate constants for the
sequential model in both branches of RC for different samples of RC
are collected in Table I.

The expressions for the electron transfer are given by the inverse
Laplace transformation. Therefore firstly we apply the Laplace
transformation to $P(t)$ in system of Egs. 16. Where the Laplace
transformation is defined as
\begin{equation}\label{}
P(s)= \int_{0}^{\infty} e^{-st}P(t)dt.
\end{equation}
Next we apply the inverse Laplace transformation of $P(s)$ where the
inverse Laplace transformation is represented by a set of simple
poles of $P(s)$. Evaluating it we obtain
\begin{equation}\label{}
P(t)= \sum_{j=1}^{5}a_{j}e^{k_{j}t},
\end{equation}
where $a_{j}$ are amplitudes and $k_{j}$ are rate kinetic constants
describing the electron transfer.

With using the model described above we would like to find kinetic
of the reaction centers of {\it Chloroflexus
aurantiacus}~\cite{Holzwarth} where on the $M$-branch the
$\mathrm{BChl}_M$ is replaced by $\mathrm{BPh}_M$ in corresponding
position. Thus {\it C. aurantiacus} RCs contain altogether three
$\mathrm{BPh}$ molecules and only one $\mathrm{BChl}$ monomer. To
characterized {\it C.aurantiacus} we start from the set of
parameters that characterize the kinetics of wild-type (WT) RCs of
Rb.sphaeroides. We use the following values of input parameters:
reorganization energies $\lambda_{ij}=800$ $cm^{-1}$, electronic
couplings $S_{ij}=0.5$ $cm^{-1}$, high frequency modes
$\omega_{ij}=1500$ $cm^{-1}$ where $i,j=1,3,5$ for $L$-side and
$i,j=1,2,4$ for $M$-side of RC. The values for electronic
couplings $V_{24}=V_{35}=32$ $cm^{-1}$,$V_{12}=V_{13}=20$
$cm^{-1}$ were used. The sink parameters
$2\Gamma_{M}/\hbar=2\Gamma_{L}/\hbar=(200ps)^{-1}$,$2\Gamma_{1}/\hbar=(170ps)^{-1}$
were used in accordance with experimental observation which
characterize the ET to quinone molecules and decay to the ground
state. Because of $\mathrm{BChl}_M$ is replaced by
$\mathrm{BPh}_M$ in corresponding position we decrease the free
energy in site $2$. The calculated rate constants and quantum
yields for the concrete energy levels are collected in Table I. We
get the following occupation probabilities for different sites:
\begin{subequations} \begin{eqnarray}
P_1(t)&=&0.05e^{-0.6t}+0.15e^{-0.37t}+0.8e^{-0.22t},\nonumber\\[1mm]
P_2(t)&=&-0.25e^{-0.6t}+0.05e^{-0.37t}+0.2e^{-0.22t},\nonumber\\[1mm]
P_3(t)&=&-0.03e^{-0.6t}-0.96e^{-0.37t}+0.99e^{-0.22t},\nonumber\\[1mm]
P_4(t)&=&0.2e^{-0.6t}-0.07e^{-0.37t}-0.44e^{-0.22t}+0.3e^{-0.01t}+0.01e^{-0.009t},\nonumber\\[1mm]
P_5(t)&=&0.02e^{-0.6t}+0.8e^{-0.37t}-1.52e^{-0.22t}-0.4e^{-0.01t}+1.1e^{-0.009t}.
\end{eqnarray}
\end{subequations}
The exponential components with very small amplitudes were neglected
in the expressions above. Time evolution of the occupation
probabilities is shown in the Fig.2.
\begin{figure}[htb]
\centering
\includegraphics[width=0.9\textwidth]{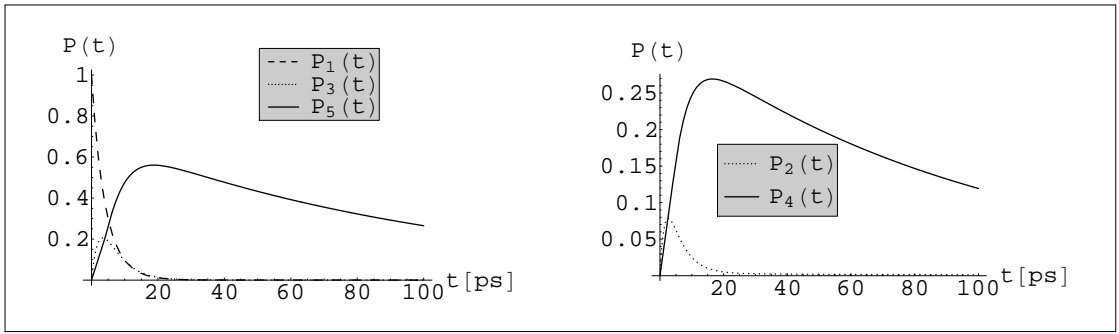}
\caption{The occupation probabilities $P(t)$ for the reaction center
{\it Chloroflexus aurantiacus} in the case if $V_{12}=V_{13}$
.}\label{kin1}
\end{figure}

We can see that in this case we get electron transfer through the M
branch, which is not in accordance with experimental observation. To
avoid this discrepancy we must assume the asymmetry in the
electronic coupling. To describe experimental kinetic of {\it
Chloroflexus aurantiacus} RC we used the following asymmetry in
electronic couplings: $V_{12}=10$ $cm^{-1}$ and $V_{13}=15$
$cm^{-1}$. We weakly decrease the coupling constants in comparison
with previous case, because of the kinetic in this RC is slower then
in the WT RC. The calculated rate constants and quantum yields for
the concrete energy levels are shown in Table I, second line. We
found
\begin{subequations} \begin{eqnarray}
P_1(t)&=&0.002e^{-0.53t}+0.018e^{-0.34t}+0.98e^{-0.12t},\nonumber\\[1mm]
P_2(t)&=&-0.05e^{-0.53t}+0.005e^{-0.34t}+0.04e^{-0.12t}+0.005e^{-0.01t},\nonumber\\[1mm]
P_3(t)&=&-0.4e^{-0.34t}+0.4e^{-0.12t},\nonumber\\[1mm]
P_4(t)&=&0.053e^{-0.53t}-0.003e^{-0.34t}-0.22e^{-0.12t}+0.13e^{-0.01t}+0.04e^{-0.009t},\nonumber\\[1mm]
P_5(t)&=&0.4e^{-0.34t}-1.3e^{-0.12t}-0.5e^{-0.01t}+1.4e^{-0.009t}.
\end{eqnarray}
\end{subequations}
In Figure 3 the behavior of the occupation probabilities
$P_{i}(t)$ is shown.
\begin{figure}[htb]
\centering
\includegraphics[width=0.9\textwidth]{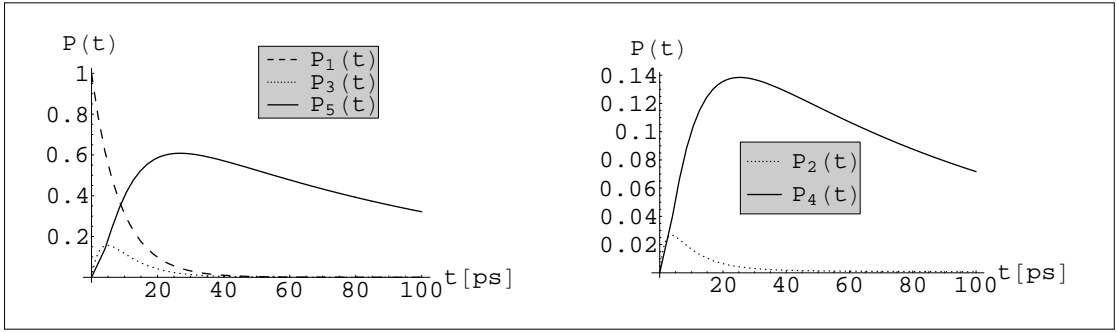}
\caption{ The occupation probabilities $P(t)$ for the reaction
center {\it Chloroflexus aurantiacus} in the case if $V_{12}\neq
V_{13}$.}\label{kin1}
\end{figure}

Now we want to elucidate the electron transfer in  {\it H(M182)L}.
In this mutant $\mathrm{BChl}_M$ is replaced with
$\mathrm{BPh}_M$. The new cofactor is referred to as $\phi_{M}$.
It is reasonable that in the {\it H(M182)L} mutant the state
$P^+\phi_{M}^-$ is lower in energy than $P^+BChl_{M}^-$ in
WT~\cite{Wood}. To explain the electron transfer in this mutant we
started from incoherent model. In this model we assume that the
energy of $P^+\phi_{M}^-$ is lower than the free energy of the
state $P^+BPh_{M}^-$ ~\cite{Holten}. The value of the free
energies used to calculate the rate constant are listed in Table
I. The occupation probabilities $P_{i}(t)$ in the case of {\it
H(M182)L} mutant are found in the form
\begin{subequations} \begin{eqnarray}
P_1(t)&=&0.33e^{-0.38t}+0.67e^{-0.25t},\nonumber\\[1mm]
P_2(t)&=&-0.02e^{-0.52t}-0.07e^{-0.38t}-0.23e^{-0.25t}+0.32e^{-0.0006t},\nonumber\\[1mm]
P_3(t)&=&-1.4e^{-0.38t}+1.4e^{-0.25t},\nonumber\\[1mm]
P_4(t)&=&0.04e^{-0.52t}-0.036e^{-0.38t}-0.04e^{-0.25t}+0.036e^{-0.0006t},\nonumber\\[1mm]
P_5(t)&=&1.1e^{-0.38t}-1.602e^{-0.25t}+0.5e^{-0.01t}+0.002e^{-0.0006t},
\end{eqnarray}
\end{subequations}
and the behavior of the occupation probabilities is shown in
Fig.4.
\begin{figure}[htb]
\centering
\includegraphics[width=0.9\textwidth]{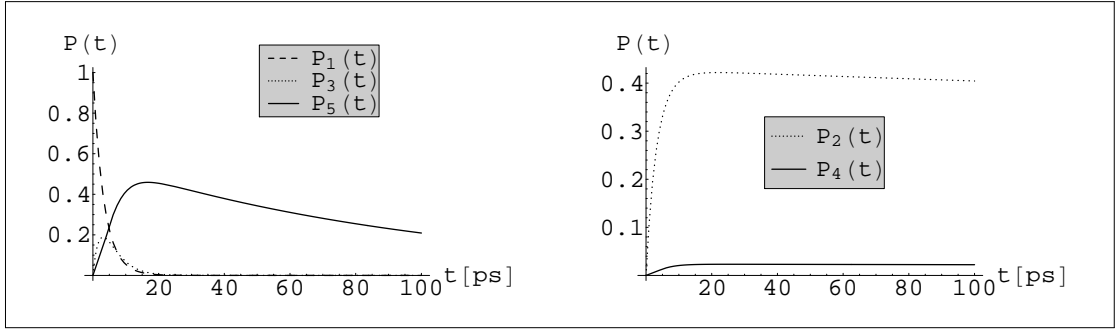}
\caption{ The occupation probabilities $P_{i}(t)$ for the mutant
{\it H(M182)L} in the case of incoherent model of electron
transfer.}\label{kin1}
\end{figure}
It was assume that the free energy of $P^+\phi_{M}^-$ is
significantly below $P^+BPh_{M}^-$ because of the electron transfer
stops at $\phi_{M}$. We get a very small probability to find
electron on $BPh_{M}$ but the quantum yields through the branch $M$
is substantial(Table I).

Now we intend to elucidate the observed ET kinetics with partially
coherent models. It means, that we assume that the reorganization
energy for ET from state $P^+\phi_{M}^-$ to state $P^+BPh_{M}^-$ is
practically zero. The electron kinetic have to be described by the
following system of equations
\begin{subequations} \begin{eqnarray}
\partial_tP_1(t)&=&-(\frac{2\Gamma_1}{\hbar}+k_{12}+k_{13})P_1(t)\nonumber\\[1mm]
& & +k_{21}P_2(t) + k_{31}P_3(t),\\[1mm]
\partial_tP_2(t)&=& -k_{21}P_2(t)- \int_0^{t}W_{24}(t-\tau)P_2(\tau)d\tau+k_{12}P_1(t)
+\int_0^{t}W_{42}(t-\tau)P_4(\tau)d\tau \\[1mm]
\partial_tP_3(t)&=&
-(k_{35}+k_{31})P_3(t)+k_{13}P_1(t)+k_{53}P_5(t),\\[1mm]
\partial_tP_4(t)&=&
-\frac{2\Gamma_M}{\hbar}P_4(t)-\int_0^{t}W_{42}(t-\tau)P_4(\tau)d\tau +\int_0^{t}W_{24}(t-\tau)P_2(\tau)d\tau,\\[1mm]
\partial_tP_5(t)&=& -(\frac{2\Gamma_L}{\hbar}+k_{53})P_5(t)+k_{35}P_3(t). \end{eqnarray}
\end{subequations}

In this case the memory function $W_{24}=W_{42} $ can be expressed
in the form: $
W_{24}(t)=2\pi\frac{|V_{24}|^{2}}{\hbar^{2}}\textrm{Re}\bigg\{\exp\bigg[-\frac{\Gamma_M+\Gamma_2}
{\hbar}t\bigg]\exp\bigg[\frac{i(\varepsilon_2-\varepsilon_4)}{\hbar}t\bigg]\bigg\}
$, where $\Gamma_{2}=0$ for our kinetic model. We now use this
partially coherent model of RC to describe {\it H(M182)L} mutation
of RC. The results of our numerical computations are collected in
Table I. We found the following expressions for occupation
probabilities $P_{i}(t)$:
\begin{subequations} \begin{eqnarray}
P_1(t)&=&0.6e^{-0.42t}+0.38e^{-0.3t}+0.02e^{-0.002t},\nonumber\\[1mm]
P_2(t)&=&-0.3e^{-0.42t}-0.201e^{-0.3t}+0.001e^{-0.009t}+0.5e^{-0.002t}+0.0004e^{-0.005t}\sin(31t),\nonumber\\[1mm]
P_3(t)&=&-1.4e^{-0.42t}+1.4e^{-0.3t},\nonumber\\[1mm]
P_4(t)&=&-0.03e^{-0.42t}-0.02e^{-0.3t}+0.03e^{-0.009t}+0.02e^{-0.002t}-0.0006e^{-0.005t}\sin(31t),\nonumber\\[1mm]
P_5(t)&=&e^{-0.42t}-1.41e^{-0.3t}+0.3e^{-0.01t}+0.01e^{-0.009t}+0.1e^{-0.002t}.\nonumber\\[1mm]
\end{eqnarray}
\end{subequations}
The behavior of the occupation probabilities $P_{i}(t)$ is shown in
Fig.5.
\begin{figure}[htb]
\centering
\includegraphics[width=0.9\textwidth]{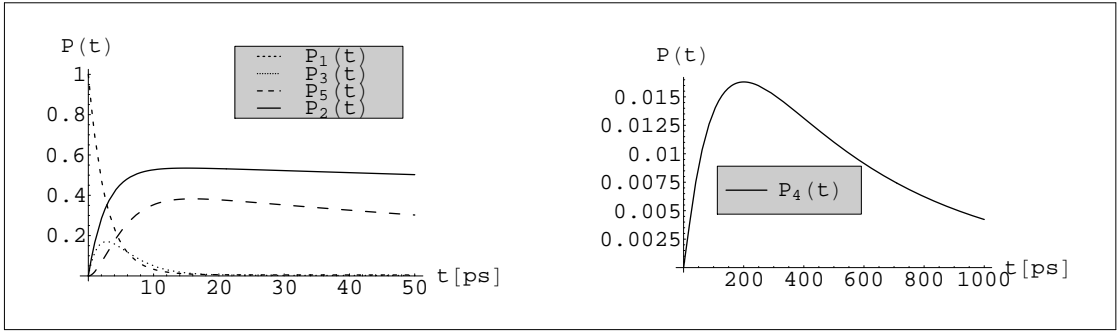}
\caption{The occupation probabilities $P(t)$ for the mutant {\it
H(M182)L} in the case of coherent model of electron transfer.
}\label{kin1}
\end{figure}
The calculated values of parameters for coherent model which
describe the mutant ${\it H(M182)L}$ RC's are collected in Table I.
\begin{table}[htb]
\begin{center}
\begin{tabular}{l c c c c c c c c c c c c c c c c}\hline\hline
\bfseries\bfseries \quad \quad &\bfseries $T$ & \bfseries
$\epsilon_{2}$ & \bfseries $ \epsilon_{3}$ & \bfseries
$\epsilon_{4}$ & \bfseries $\epsilon_{5}$ & \bfseries $1/k_{12}$&
\bfseries $1/k_{21}$ & \bfseries $1/k_{13}$ & \bfseries $1/k_{31}$ &
\bfseries $1/k_{24}$ & \bfseries $1/k_{42}$ &
&&\\
\bfseries $Sample$&$K$&$cm^{-1}$&$cm^{-1}$&$cm^{-1}$&$cm^{-1}$&$ps$&$ps$&$ps$&$ps$&$ps$&$ps$&$\Phi_{G}$ &$\Phi_{M}$&$\Phi_{L}$\\
\hline\bfseries ${\it C. aurant.}$
&295&-50&-450&-1000&-2000&12&15&6&52&2&188&0.02&0.31&0.67\\
\bfseries $V_{12}=V_{13}$&200&&&&&15&21&5&138&2&1559&0.02&0.25&0.73\\
\hline\bfseries ${\it C. aurant.}$
&295&-50&-450&-1000&-2000&47&60&11&93&2&188&0.05&0.16&0.79\\
\bfseries $V_{12}\neq V_{13}$&200&&&&&58&84&10&246&2&1559&0.05&0.12&0.83\\
\hline\bfseries ${\it H(M182)L}$
&295&-1600&-450&-1000&-2000&8&10005&6&52&37&2.1&0.02&0.41&0.57\\
\bfseries $Incoherent$&200&&&&&9&443897&5&138&132&1.8&0.02&0.37&0.61\\
\hline\bfseries ${\it H(M182)L}$
&295&-850&-450&-1000&-2000&5&292&6&52&4394&4394&0.03&0.12&0.85\\
\bfseries $Coherent$&200&&&&&4&1867&5&138&4394&4394&0.02&0.38&0.6\\
\hline\hline
\end{tabular}
\end{center}
\caption{{\footnotesize The computed rate constant $1/k_{ij}$ and
quantum yields dependent on temperature for reaction centers and
some mutants of RC's.} The rate constants $1/k_{35}=3(4)$$ps$ for
$T=295(200)K$ and $1/k_{53}=5166(264544)$$ps$ for $T=295(200)K$ are
the same for all RC's and mutations describing in the Table I.}
\label{tab2}
\end{table}

\newpage
\section{Conclusion}
We have dealing with electron transfer in the reaction center of
{\it Chloroflexus aurantiacus} and  {\it Rhodobacter sphaeroides}
{\it H(M182)L} mutated RC. In spite of their structural similarity,
the functionality is very different. {\it H(M182)L} mutant reveal
the M brunch active in electron transfer. In the previous
papers~\cite{Plato,Michel,Pudlak,Pincak,Pudlak1} were discussions
about what is dominant factor which causes the asymmetry in the
electron transfer. At the beginning it was assumed that the
asymmetry in the coupling parameters is dominant. The later the
experimental work brought a doubt about dominance of electron
coupling as a mechanisms which cause the asymmetry in ET trough
branches~\cite{Wood,Holten}. Now, it is assume that the asymmetry in
energetics also contribute to asymmetry of ET through M and L
branches. We have showed that in the {\it Chloroflexus aurantiacus}
RC we must have minimally 2:3 ratio of the electron transfer
integrals for  $P^*\phi_{M}\leftrightarrow P^+\phi_{M}^- $ and  for
$P^*BChl_{L}\leftrightarrow P^+BChl_{L}^-$ to explain the observed
ET kinetics. In the case of {\it Rhodobacter sphaeroides} {\it
H(M182)L} mutated RC we used two models to describe the ET. In
incoherent model we must use a very low free energy of the
$P^+\phi_{M}^- $ state in comparison to the free energy of
$P^+BPh_{M}^- $ state to get a small probability to find electron on
the $BPh_M$ molecule. Despite this assumption we get relatively
strong outlet through the M branch. In the partially coherent model
we assume that the ET between molecules $\phi_M$ and $BPh_M$ have
coherent character, this means that ET is so fast that the bath does
not have sufficient time to relax to the new thermal equilibrium
before the particle moves away. The result is outlet trough the M
branch which is in accordance with experimental observations. The
problem of both coherent and incoherent models is that predict not
enough decay to the basic state. The similar model ought to be used
to characterized electron transfer in PSII (PSI) system, where dimer
molecules of $BChl$ are not so close as in the bacterial reaction
centers. The temperature dependence of the kinetics and QYs was also
computed. We can see that there are differences in dependence of QY
on the temperature in both models which were assumed.

\vskip 0.2cm \vskip 0.2cm The work was supported in part by VEGA
grant 2/7056/27. of the Slovak Academy of Sciences, by the Science
and Technology Assistance Agency under contract No. APVT-51-027904.

\newpage

\end{document}